\documentclass[twocolumn,secnumarabic,amssymb, nobibnotes, aps, prd, superscriptaddress,floatfix]{revtex4-2}

\newcommand{\comp}{\tilde{c}}
\newcommand{\fNN}{f^{(1)}}
\newcommand{\fNNN}{f^{(2)}}

\usepackage[normalem]{ulem}
\usepackage{xcolor}

\usepackage{lipsum}
\usepackage{graphicx}
\usepackage{amsmath}
\begin{document}

\title{Spinodal decomposition and domain coarsening in a multi-layer Cahn-Hilliard model
for lithium intercalation in graphite}%

\author{Antoine Cordoba}%
\affiliation{Univ. Grenoble Alpes, CEA, LITEN, DEHT, 38054 Grenoble, France}
\affiliation{Laboratoire de Physique de la Mati\`ere Condens\'ee, CNRS, Ecole Polytechnique, 
Institut Polytechnique de Paris, 91120 Palaiseau, France}
\author{Marion Chandesris}%
\affiliation{Univ. Grenoble Alpes, CEA, LITEN, DEHT, 38054 Grenoble, France}
\author{Mathis Plapp}%
\email[M. Plapp: ]{mathis.plapp@polytechnique.fr}
\affiliation{Laboratoire de Physique de la Mati\`ere Condens\'ee, CNRS, Ecole Polytechnique, 
Institut Polytechnique de Paris, 91120 Palaiseau, France}
\date{\today}%

\begin{abstract}
During the intercalation of lithium in layered host materials such as graphite, lithium atoms
can move within the plane between two neighboring graphene sheets, but cannot cross the sheets. 
Repulsive interactions between atoms in different layers lead to the existence of ordered phases 
called ``stages'', with stage $n$ consisting of one filled layer out of $n$, the others being empty.
Such systems can be conveniently described by a multi-layer Cahn-Hilliard model, which can be seen
as a mean-field approximation of a lattice-gas model with intra- and interlayer interactions between
lithium atoms. In this paper, the dynamics of stage formation after a rapid quench to lower temperature
is analyzed, both by a linear stability analysis and by numerical simulation of the full equations.
In particular, the competition between stages 2 and 3 is studied in detail. The linear stability
analysis predicts that stage 2 always grows the fastest, even in the composition range where stage 3
is the stable equilibrium state. This is borne out by the numerical simulations, which show that
stage 3 emerges only during the non-linear coarsening of stage 2. Some consequences of this finding
for the charge-discharge dynamics of electrodes in batteries are briefly discussed.
\end{abstract}

\maketitle

\section{Introduction}

The advent of portable consumer electronics in the 1970s has boosted the research on lithium batteries, leading in 1991 to the first commercialization by Sony and Asahi Kasei of a rechargeable lithium-ion battery (LIB), with graphite at the negative electrode and LiCoO$_2$ at the positive electrode~\cite{Scrosati2011,Xie2020}.
Beyond the portable electronic devices market, LIBs are now successfully used for electromobility and are considered as a technology of interest for stationary energy storage~\cite{Dunn2011}.
Current research investigates a large variety of LIB active materials and postlithium ion concepts to meet the demand for better performing and safer storage technologies~\cite{Armand2020}. Nevertheless, graphite still remains the dominant negative electrode material in commercial cells~\cite{Andersen2021}.
It has the benefit of being abundant and cheap, to have an excellent cycling capacity, a low thermodynamic potential and a decent specific capacity of 372 mAh/g.
However, it also has some well-known limitations, such as its slow kinetics, solid-electrolyte interface (SEI) formation upon cycling, and lithium plating upon fast-charge~\cite{Asenbauer2020}. A better understanding of its intrinsic properties is necessary to overcome these problems.

Graphite is a lamellar material made of two-dimensional graphene layers stacked upon each other and linked by relatively weak van der Waals forces.
Lithium ions can be reversibly intercalated in between adjacent graphene layers~\cite{Dahn1991,Ohzuku1993a,Billaud1996}, but the diffusion of lithium across the graphene layer is highly unlikely~\cite{Meunier_2002, Yao2012, Krishan2013, Li2021}.
For high enough lithium concentration, a staging phenomenon occurs: lithium ions form ordered structures with filled layers regularly separated by empty ones~\cite{Rudorff1938}. These structures are referred to by a stage number, stage \(n\) meaning that one out of every \(n\) galleries is filled, with the others being empty. The different stages correspond to ordered phases of the layered structure, and coexistence between different stages is possible. During charge and discharge, a sequence of stages occurs, and the voltage-charge curve exhibits plateaux during the replacement of one stage by another one.

There is a large consensus with regards to the structures of lithiated graphite at high lithium concentrations~\cite{Guerard1975,Billaud1981}. In contrast, for low lithium concentrations, i.e, stages with a number higher than 2, there is an ongoing discussion about the precise compositions and structure of the stages~\cite{Dahn1991,Ohzuku1993a,Billaud1996}. Moreover, a hysteresis between charge and discharge is also reported both in the entropy measurement~\cite{Allart2018} and in the structural evolution between the stages~\cite{Hess2013a,Yao2019Gr,Didier2020,Schmitt2021}. Indeed, during lithiation the sequence is pure graphite, dilute stage 1, stage 3L, stage 2 and stage 1  whereas during delithiation a stage 2L can be observed between stage 2 and 3L. Here, the denomination L stands for ``liquid-like'', in reference to the absence of lithium ordering in the plane between the graphene layers. For a good description of these transitions, a model is needed that captures all the different stages with their respective symmetries, in particular stage 3.

\begin{figure}[ht!]
    \centering
    \includegraphics[width=\linewidth]{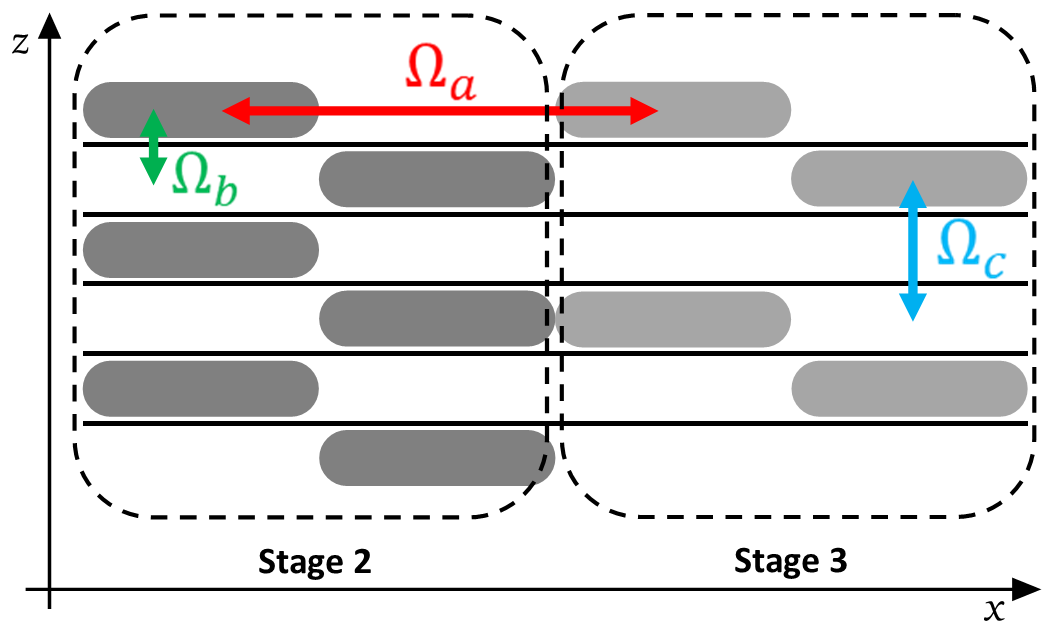}
    \caption{Illustration of graphite staging structure based on the Daumas-Hérold representation. Black lines represent graphene planes and grey domains the lithium islands. Stage 2 is represented on the left and stage 3 on the right together with the interactions between lithium atoms.}
    \label{fig:layers}
\end{figure}

The representation in which a layer is completely full or empty, known as the Rüddorf-Hofmann model~\cite{Rudorff1938}, cannot explain the kinetics of staging transitions between stages with even and odd numers, for example between stages 2 and 3.
This is because, as lithium migration from one layer to another can occur only at the edges of the graphite particles or at grain boundaries, the Rüddorf-Hofmann representation would require an entire gallery to be emptied or filled for this transition to occur.
Five decades ago, Daumas and Herold \cite{Daumas1969} proposed the domain model, where the lithium ions occupy the galleries as islands, locally maintaining the staging. This structure in islands is schematically illustrated in Fig.~\ref{fig:layers} with a coexistence of two regions, one in stage 2 on the left and one in stage 3 on the right.
The fact that graphite tends to phase-separate into these Li-rich and Li-poor regions is a key element to understand the kinetics of stage transitions and will affect both the lithium transport inside the graphite particles and the lithium insertion at its surface~\cite{Bazant2013,smith_intercalation_2017}.
A model able to capture the formation of these Li-rich and Li-poor domains is therefore necessary to study the dynamics of the transition between stage 2 and 3.

Diffuse-interface models based on the physics of first-order phase transitions~\cite{Cahn1958,Cahn1959,Cahn1959b} have been widely used to study phase-separating systems~\cite{Hohenberg77,LangerBegRohu}. They naturally capture domain formation and interface motion, and work on diffusive time scales.
In the field of LIBs, renewed attention has been given to these approaches for their ability to predict and simulate phase separation in two-phase materials such as LFP~\cite{Han2004,Singh2008,Cogswell2012,Li2018a,Katrasnik2019,Daubner2021}.
To capture staging in intercalation compounds, a multi-layer diffuse-interface model has been developed in the 1980's~\cite{Safran1980,Millman1983}.
In this approach, the concentration of lithium in each gallery is treated by a separate evolution equation.
The model considers attractive in-plane interactions and repulsive interplanar interactions between intercalants.
A mean-field approximation is then used to obtain a free-energy model. It contains a square-gradient energy term for each layer that arises from the attractive in-plane interactions.

This multi-layer diffuse-interface framework is an attractive tool to analyze multi-phase lamellar LIB active materials such as graphite.
Hawrylak and Subbaswamy~\cite{hawrylak_kinetic_1984} simulated a six-layer system and demonstrated the intrinsic tendency of the system to form Daumas-Herold islands and its ability to transition between stages 2 and 3. 
More recently, using a bi-layer model, Bazant \textit{et al.}~\cite{Fergusson2014,guo_li_2016,smith_intercalation_2017} have studied the dynamics of intercalation between three stable phases, stage 1' (dilute solid solution), stage 2 and stage 1 and compared it to the direct optical imaging from Guo \textit{et al.} \cite{guo_li_2016}.
They highlight the occurrence of "checkerboard" domains which cannot be captured by solid-solution models based on Fickian diffusion. 
Such models can be refined in various ways to yield a more quantitative description of specific systems. For instance, the free-energy model of Hawrylak and Subbaswamy~\cite{hawrylak_kinetic_1984}
leads to the formation of a stage \(3/2\) which has two full layers and one empty and is thus symmetric to the stage 3 (through an exchange of full and empty layers). Such a stage is, however, not observed in the experiments on graphite. Chandesris {\em et al.}~\cite{chandesris_thermodynamics_2019}, showed that the asymmetry of the graphite phase diagram can be captured by introducing a continuous screening effect for the repulsive interaction between second neighbours. The resulting model predicts the formation of the staged phases 2 and 3, without the occurrence of the 3/2 stage. Agrawal and Bai~\cite{Agrawal2023} used the two-layer model of Bazant \textit{et al.}~\cite{Fergusson2014,smith_intercalation_2017} in a multi-scale simulation framework to describe charge and discharge of a composite electrode containing many graphite particles.
Given the computational cost associated to the multi-layered phase-field approach, some authors~\cite{smith_intercalation_2017,Thomas-Alyea2017,Ahn2023,Lian2023,lu2023multiscale} also developed reduced phase-field models to capture some of the features of the phase transforming behavior of lithiated graphite while studying the coupling with the other transport phenomena at the electrode scale and the consequences during fast charge.

The latter references go beyond a simple use of multilayer phase-field models. Indeed, the electrode of a real battery has a complex architecture, and different transport processes occur in series during the charge-discharge process: lithium atoms have to be transported through the electrolyte, then enter the active material, which can have itself a complex and polycrystalline microstructure, through an intercalation reaction at the surface, and finally migrate inside the particles of the active material. Each of these transport processes occurs on a different scale and exhibits different complexities, so that multiscale and multiphysics modelling are required for a holistic modeling of battery operation \cite{Thomas-Alyea2017,Danner2016,Franco2019,Taghikhani2020,Ahn2023,Lian2023,lu2023multiscale}. However, a more detailed knowledge on each of the sub-processes is also needed for a precise description. 

Whereas the transition between stages 2 and 1 at high lithium concentration is certainly more relevant for the optimization of battery electrode operation in the fast charging regime \cite{lu2023multiscale}, the competition between stages 2 and 3 at low lithium concentration remains a poorly understood problem that is interesting from a fundamental point of view and might impact other aging mechanisms induced by internal mechanical stress and concentration dependent side reactions. To shed new light on this question, we deliberately choose to focus on a particularly simple situation: the spinodal decomposition of a homogeneous unstable initial state, without external driving, in a one-dimensional stack of graphite without defects or grain boundaries. The great advantage of this setting is that it can by analyzed analytically by a linear stability analysis. Hawrylak and Subbaswamy~\cite{hawrylak_kinetic_1984} mention such an analysis, but do not provide any details. Agrawal and Bai~\cite{Agrawal2023} perform a linear stability analysis in the context of open driven systems, taking into account local interfacial currents. But
since the underlying model has only two-layers, the growth rate of the stage 3 inside the particle cannot be captured. In conclusion, no complete treatment of this question is as of yet available.

Here, we lay out a general formulation for the stability analysis of multi-layer staging models without external driving, and apply it to the model of Chandesris {\it et al.}~\cite{chandesris_thermodynamics_2019}. The analysis is carried out in terms of normal modes, so that any stage that is compatible with the symmetries of the system can appear. Therefore, competition between stages 2 and 3 can be studied in systems with a number of layers that is a common multiplier of 2 and 3; we choose the smallest such number, which is six. We confirm a statement already made by Hawrylak and Subbaswamy~\cite{hawrylak_kinetic_1984}, namely, that stage 2 grows faster than stage 3, even for initial compositions for which stage 3 is the thermodynamically stable state. We conduct a systematic parametric study to ensure that this result is generic to all models that exhibit the experimentally observed staging sequence. Numerical simulations are then carried out to verify this prediction, and to study how the system approaches the correct equilibrium by nonlinear domain-coarsening processes.

In the following, we will first introduce the model in Sec. \ref{sec:model}, and then present the linear stability analysis in Sec.~\ref{sec:theory}. In Sec.~\ref{sec:num}, numerical simulations are performed to test the prediction on the LSA and to study the transition from stage 2 to stage 3 during domain coarsening. Some possible consequences of these findings for behavior during charge and discharge are discusses in Sec.~\ref{sec:discussion}, followed by the general conclusion.

\section{Model\label{sec:model}}
We consider a system of $N_z$ graphene layers stacked along the $z$-axis (See Fig.~\ref{fig:layers}). 
Since lithium atoms cannot jump between layers, a separate concentration field is defined in each layer. These fields are functions of the continuous coordinates in the plane $x$ and $y$~\cite{hawrylak_kinetic_1984,smith_intercalation_2017}.

\subsection{Free-energy model~\label{sec:free-energy}}
Let $c_j(x,y)$ be the local lithium concentration (in units of mole per unit volume) in layer number $j$.
We define the scaled concentration $\tilde{c}_j(x,y)=c_j(x,y)/c_{\rm max}$, where $c_{\rm max}$ is the maximum possible concentration in the layer for the studied material. The latter can be related to the number of possible intercalation sites per unit volume, $N_V$, by $N_V=\mathcal{N}_A c_{\rm max}$, where $\mathcal{N}_A$ is the Avogadro number.
The total Gibbs free energy of the system, $G$, may be written in terms of local volumetric free energy density $\mathcal{G}$, as: 
\begin{align}
    \label{eq:FreeNRJ}
    G & =\int_V \mathcal{G}~dV\notag \\ 
    & = \sum_{j=1}^{N_z} L_j \int_S \bigg[ \dfrac{1}{2}\kappa(\nabla\comp_j)^2+g(\comp_j)\notag \\
    &+g^{(1)}(\comp_j,\comp_{j\pm 1})
    +g^{(2)}(\comp_j,\comp_{j\pm 1},\comp_{j\pm 2})\bigg] dx\,dy
\end{align}
Here, $V$ is the volume of the material, and $S$ and $L_j$ are the surface area and the thickness of one graphite layer. In this work, we will neglect the volume dilatation due to intercalation and take $L_j=L_z$ as a constant independent of $j$.
In the first term of Eq.~(\ref{eq:FreeNRJ}), the gradient operator is applied in the plane (along directions $x$ and $y$), and \(\kappa\) is a constant coefficient of dimension energy per unit length. This is the gradient-square term introduced by Cahn and Hilliard~\cite{Cahn1958}. It describes the energy penalty linked to any inhomogeneity in concentration and arises from attractive in-plane interactions between the lithium atoms.

The second term in Eq.~(\ref{eq:FreeNRJ}) describes the thermodynamics of a single layer.
The simplest model that leads to phase separation within a layer is the regular solution model that is a mean-field approximation of the lattice gas model~\cite{LangerBegRohu}:
\begin{align}
    \label{eq:NRJ_intra}
    g(\comp_j)=&N_V\bigg\{k_BT\big[\comp_j\ln(\comp_j)+(1-\comp_j)\ln(1-\comp_j)\big]\notag \\
    &+\Omega_a\comp_j(1-\comp_j)+\comp_j\mu_{\rm ref}\bigg\}
\end{align}
where \(k_B\) is the Boltzmann constant, and \(T\) is the temperature.
The first term in Eq.~(\ref{eq:NRJ_intra}) is due to the configurational entropy of full and empty intercalation sites, while the second term results from the attractive interaction between lithium atoms within the same plane, with \(\Omega_a\) being the corresponding interaction parameter.
Finally, \(\mu_{\rm ref}\) is a reference potential.

The last two terms of Eq.~(\ref{eq:FreeNRJ}) arise from the interactions between lithium atoms in different planes. The first-neighbor interaction energy $g^{(1)}$ may be written as: 
\begin{align}
    \label{eq:NRJ_inter_NN}
    g^{(1)}(\comp_j,\comp_{j\pm 1})=N_V\dfrac{\Omega_b}{2}(\comp_j\comp_{j+1}+\comp_j\comp_{j-1})
\end{align}
where \(\Omega_b\) describes the interaction strength. Since this parameter is positive, it will cost energy to have high lithium concentration in adjacent layers for a given position \((x,y)\), which will favor the appearance of stage 2.

Interlayer interactions of longer range are necessary to obtain staging phenomena beyond stage 2.
A simple repulsive second-neighbor interaction of the form $g^{(2)}=\Omega_c \comp_j\comp_{j+2}$,
as in~\cite{Safran1980,hawrylak_kinetic_1984} would favor stage 3.
However, it was found that this expression leads to a phase diagram that is symmetric with respect to half-filling ($\tilde c_j=1/2$): a new stage 3/2 appears, in which two full layers are separated by a single empty one~\cite{Millman1983,chandesris_thermodynamics_2019}. Since this stage is not observed in experiments, we therefore use a more complex expression for the second-neighbor interactions which includes a continuous screening effect introduced in a previous work~\cite{chandesris_thermodynamics_2019}:
\begin{align}
    \label{eq:NRJ_inter_NNN}
    g^{(2)}(\comp_j,\comp_{j\pm 1},\comp_{j\pm 2}) = &N_V\dfrac{\Omega_c}{2}\bigg[\comp_j(1-\comp_{j+1})\comp_{j+2}\notag \\
    &+\comp_j(1-\comp_{j-1})\comp_{j-2}\bigg]
\end{align}
With this expression, high concentrations of intercalants in the next-nearest neighbor layers are energetically penalized, except if intercalants are also present in between.
The effect of this screening can be assessed considering two limits:
\begin{itemize}
    \item $\Omega_c(1-\comp_{j+1}) \approx \Omega_c$ for low value of $\comp_{j+1}$, favoring the formation of stage 3 at low lithium concentration;
    \item $\Omega_c(1-\comp_{j+1}) \approx 0$ for high value of $\comp_{j+1}$, favoring the formation of stage 2 (instead of stage 3/2) at high lithium concentration.
\end{itemize}
Interactions of longer range are mentioned in several papers~\cite{Safran1980,Millman1983,hawrylak_kinetic_1984}.
As these interactions rapidly decrease with the distance between layers and are not crucial to analyze transitions between stage 2 and stage 3, they are neglected in the present study. 

\subsection{Kinetics}
The kinetic equations for the concentration fields are formulated using the thermodynamic theory of irreversible processes~\cite{Cahn1958, Groot1962, Kawasaki1966, Hohenberg_1977}. Since the atoms cannot jump from one layer to another, each concentration field satisfies a conservation law,
\begin{align}
    \label{eq:mass_conservation}
    \dfrac{\partial\comp_j}{\partial t}=-\nabla \cdot J_j,
\end{align}
where \(J_j\) is the flux of lithium atoms in layer \(j\). This flux can be expressed as 
\begin{align}
    \label{eq:flow}
    J_j=-\mathcal{M}\comp_j\nabla \mu_j,
\end{align}
where \(\mu_j\) is the chemical potential in layer \(j\).
The latter is defined as the variational derivative of the free energy with respect to the concentration within this layer: 
\begin{align}
    \label{eq:potential}
    \mu_j=\dfrac{1}{c_{max}}\dfrac{\delta G}{\delta \comp_j}
\end{align}
In Eq.~(\ref{eq:flow}), \(\mathcal{M}\) is the mobility. Whereas a constant mobility is used in some papers \cite{hawrylak_kinetic_1984,chandesris_thermodynamics_2019}, we choose here an expression that is consistent with diffusion of atoms on a lattice~\cite{smith_intercalation_2017},
\begin{align}
    \label{eq:mobility}
    \mathcal{M}(\comp_j)=\dfrac{D}{\mathcal{N}_Ak_BT}(1-\comp_j),
\end{align}
with $D$ being the tracer diffusivity in the dilute limit.
Gathering Eq.~(\ref{eq:mass_conservation}) to (\ref{eq:mobility}), we have: 
\begin{align}
    \label{eq:Cahn_Hilliard}
    \dfrac{\partial \comp_j}{\partial t}=\nabla\cdot \bigg( M(\comp_j)\nabla\dfrac{\delta G}{\delta\comp_j}\bigg)
\end{align}
where we have defined 
\begin{align}
    \label{eq:mobilite}
    M(\comp_j)=\dfrac{D}{N_Vk_BT}\comp_j(1-\comp_j).
\end{align}
The factor \(\comp_j(1-\comp_j)\) ensures that there is no mass transport when there is no lithium, and when the layer is completely full.

\section{Theoretical analysis\label{sec:theory}}
\subsection{Phase diagram\label{sec:phase-diag}}
For given interaction parameters, the phase diagram can be determined by minimization of the free-energy functional under the constraint of mass conservation. The details of this procedure are described in Ref.~\cite{chandesris_thermodynamics_2019}; therefore, we will summarize here only the most important points.

For a single layer, the intra-layer free energy $g(\tilde c_j)$ is a double-well function due to the competition between configurational entropy and the attractive in-plane interactions of strength $\Omega_a$. This leads to the coexistence of concentrated and dilute domains in the same layer, with equilibrium compositions given by the common tangent to $g(\tilde c_j)$~\cite{smith_intercalation_2017,chandesris_thermodynamics_2019}.
For multiple layers, the formation of the ordered stages is driven by the repulsive interlayer interactions  (illustrated by $\Omega_b$ and $\Omega_c$ in Fig.~\ref{fig:layers}). A given stage at equilibrium is characterized by the co-existence of lithium-poor and lithium-rich domains. Since all the layers are in contact with the same external medium (the electrolyte), the chemical potential must be the same in all layers. Therefore, the equilibrium compositions of a given stage can again be determined by a double-tangent construction. This also yields the equilibrium potential, which can be different for each stage. The equilibrium compositions also differ from the ones in a single layer.

The equilibrium between two different stages occurs when their chemical potential and their grand potential, defined as \(\mathcal{G}(c)-\mu^{eq}c\), are the same. The coexisting stages have different average compositions (filling fractions). For filling fractions that fall in between these two values, the two stages coexist, with volume fractions that are determined by the lever rule. 

\begin{figure}
    \centering
    \includegraphics[width=\linewidth]{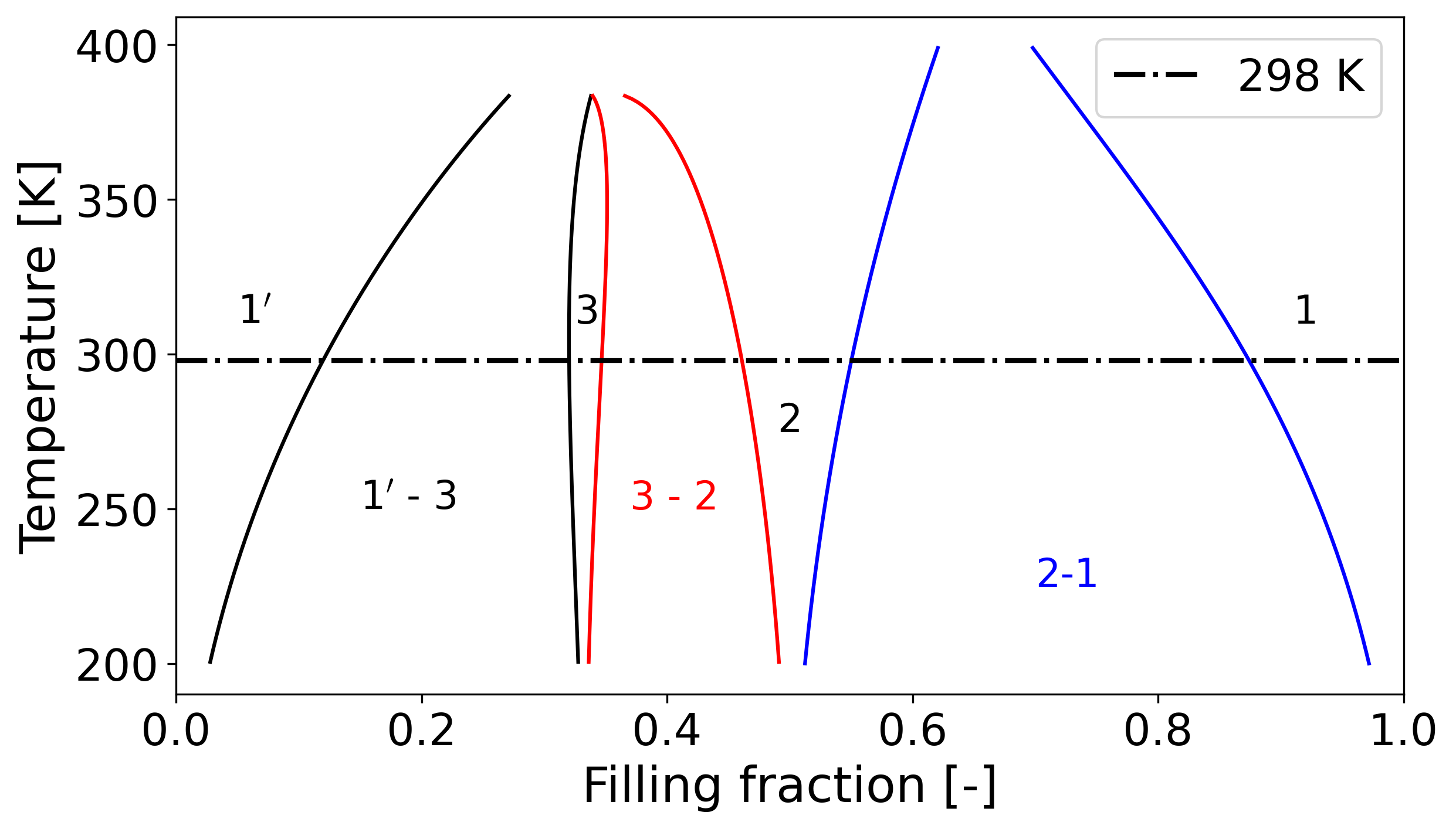}
    \caption{Phase diagram of lithium intercalated graphite with the thermodynamic parameters of TABLE~\ref{tab:phase_diagram}.}
    \label{fig:Phase_Diagram}
\end{figure}

In the following, we will use a parameter set that was developed in Ref.~\cite{chandesris_thermodynamics_2019}, see Table~\ref{tab:phase_diagram}.
The corresponding phase diagram is presented in Fig.~\ref{fig:Phase_Diagram}.
With the screened second neighbor interaction of Eq.~(\ref{eq:NRJ_inter_NNN}), the phase diagram is not symmetric. For mean filling fractions below 0.5, there are two regions of phase coexistence: one for dilute stage 1/stage 3 and one for stage 3/stage 2. In contrast, for mean filling fractions above 0.5, the only phase coexistence is between stage 2 and stage 1.
At room temperature, \(298~K\), the sequence of stages that occurs for increasing filling fraction is: dilute stage \(1'\), coexistence of stages \(1'\) and \(3\), pure stage \(3\), coexistence of stages \(3\) and \(2\), pure stage \(2\), coexistence of stages \(2\) and \(1\), and finally pure stage \(1\). When only one stage is present, the equilibrium chemical potential is an increasing function of the filling fraction; in contrast, during coexistence of two stages, the potential remains fixed to its equilibrium value at phase coexistence, and only the phase fractions change. Therefore, the curve of potential versus filling fraction exhibits a plateau whenever two stages coexist. Note that the quantity that is usually measured is the electric potential, which is proportional to the negative of the chemical potential.
In the model, the value of $\Omega_a$ determines the concentration difference between dense and dilute domains and the width of the voltage plateaus, whereas the interlayer interaction energies \(\Omega_b\) and \(\Omega_c\) will influence the potential differences between the different plateaus and the limits of phase coexistence. 
The orders of magnitude of all the interaction parameters are similar to previously reported values~\cite{Han2004,Fergusson2014,smith_intercalation_2017}.

The maximum concentration of lithium \(c_{\rm max}\) can be evaluated theoretically from the crystal structure or estimated experimentally from the reversible capacity.
The value used in this work, $c_{\rm max}=30\,000$ mol/m$^3$, falls within the range of averaged values reported in the liiondb database for graphite~\cite{Wang2021}. The value of the reference potential \(\mu_{\rm ref}\) was set to zero as it has no impact on the phase diagram. 
The values of the gradient penalty coefficient and of the diffusivity will be discussed later. 

\begin{table}[ht!]
    \begin{tabular}{c|c}
        \hline
        \(c_{\rm max}\) & \(30~000\quad \rm{mol}\; \rm{m}^{-3}\) \\
        \(\Omega_a\) & \(64.3\quad \rm{meV}\) \\
        \(\Omega_b\) & \(23.1\quad \rm{meV}\) \\
        \(\Omega_c\) & \(4.1\quad \rm{meV}\) \\
        \(\mu_{ref}\) & \(0\quad \rm{J}\) \\
        \(\kappa\) & \(3\times 10^{-6}\quad \rm{J}\; \rm{m}^{-1}\)\\
        \(D\) & \(1.25\times 10^{-12} \quad \rm{m}^2\; \rm{s}^{-1}\)\\
        \hline
    \end{tabular}\\
\caption{List of model parameters.\label{tab:phase_diagram}}
\end{table}

Compared to experimental phase diagrams close to room-temperature~\cite{Fischer1983,GIC1990,Dahn1991}, the predicted diagram does not capture all the details at low average filling fraction and overestimates the mean filling fraction of stage 3.
This might be improved by including longer-range lithium interactions~\cite{Millman1983} and/or more complex interactions with the graphite structure~\cite{Rykner2022}.
These more complex approaches, however, still need to demonstrate their ability to fully describe the transitions occurring at low filling fraction.

\subsection{Spinodal decomposition}
\label{sect::Growth_rate}
Spinodal decomposition is a spontaneous transition from a single homogeneous but unstable phase into two phases with lower free energy. It can be induced by a quench, that is, the system temperature is rapidly changed. Even though such experiments have, to our knowledge, not been reported for intercalation materials, it is still instructive to analyze this situation. Indeed, the initial stages of spinodal decomposition can be understood analytically with the help of a linear stability analysis.

To conduct such an analysis for our model, we consider a system with periodic boundary conditions in the $z$-direction. Furthermore, in the following we simplify the calculations by supposing that the concentration is invariant along the $y$ direction; the layers are thus one-dimensional. This represents a cut through a graphite particle in a direction that is normal to the particle-electrolyte interface and to the graphene planes. The results of the stability analysis would be unchanged for a full two-dimensional treatment because of the rotational invariance within the planes. A particle of linear size $L$ is considered, with no-flux boundary conditions at both sides.

To make notations more compact, the expressions of the interaction energies are rewritten as follows:
\begin{align}
    g^{(1)}(\comp_j,\comp_{j\pm 1})=\dfrac{1}{2}\bigg[\fNN(\comp_j,\comp_{j+1})+\fNN(\comp_j,\comp_{j-1})\bigg]
\end{align}
with \(\fNN(\comp_j,\comp_{j+1})=N_V\Omega_b\comp_j\comp_{j+1}\). Similarly, for the next-nearest neighbor interaction term, we write:
\begin{align}
    g^{(2)}(\comp_j,\comp_{j\pm 1},\comp_{j\pm 2})=\dfrac{1}{2}&\bigg[\fNNN(\comp_j,\comp_{j+1},\comp_{j+2})\notag\\
    &+\fNNN(\comp_j,\comp_{j-1},\comp_{j-2})\bigg]
\end{align}
with \(\fNNN(\comp_j,\comp_{j+1},\comp_{j+2})=N_V\Omega_c\comp_j(1-\comp_{j+1})\comp_{j+1}\).
Furthermore, in the following, \(\partial_{\ell mn}f\) corresponds to the derivation of \(f\) \(\ell\) times with respect to its first argument, \(m\) times with respect to its second argument and \(n\) times with respect to the third one, and \(\partial_t\) is the regular time derivative.

With these notations, Eq.~(\ref{eq:Cahn_Hilliard}) combined with the free energy model of Eqs.~(\ref{eq:FreeNRJ}) to~(\ref{eq:NRJ_inter_NNN}) reads: 
\begin{align}
    \partial_t&\comp_j=\nabla\cdot M\nabla\bigg[-\kappa\nabla^2\comp_j+g'(\comp_j)+\partial_{10}\fNN(\comp_j,\comp_{j+1})\notag \\
    &+\partial_{10}\fNN(\comp_j,\comp_{j-1})+\partial_{100}\fNNN(\comp_j,\comp_{j+1},\comp_{j+2})\notag \\
    &+\partial_{100}\fNNN(\comp_j,\comp_{j-1},\comp_{j-2})+\partial_{010}\fNNN(\comp_{j-1},\comp_j,\comp_{j+1})\bigg]
    \label{eq:CH+Free_NRJ}
\end{align}
To linearize this expression, we consider that all filling fractions consist of fluctuations around an average initial composition \(\bar{c}\) that is the same for all layers:
\begin{align}
    \label{eq:comp_linear}
    \comp_j(x,t)=\overline{c}+\delta\comp_j(x,t)
\end{align}
Using Taylor series and neglecting all term that are non-linear in the perturbations (\(\delta\comp_j \delta\comp_i\) etc.), the following system is obtained:  
\begin{align}
    \label{eq:CH+Free_NRJ_lin}
    \partial_t&\delta\comp_j=\nabla\cdot M\nabla\bigg[-\kappa\nabla^2\delta\comp_j +g''\delta\comp_j\notag\\
    &+\partial_{11}\fNN\delta\comp_{j+1}+2\partial_{20}\fNN\delta\comp_j+\partial_{11}\fNN\delta\comp_{j-1}\notag\\
    &+\partial_{101}\fNNN\delta\comp_{j+2}+2\partial_{110}\fNNN\delta\comp_{j+1}+2\partial_{200}\fNNN\delta\comp_j\notag\\
    &+\partial_{101}\fNNN\delta\comp_{j-2}+2\partial_{110}\fNNN\delta\comp_{j-1}+2\partial_{020}\fNNN\delta\comp_j\bigg]
\end{align}
where only the zeroth order term is kept in the mobility, and all the derivatives with respect to $\tilde c$ are evaluated at the unperturbed concentration $\bar c$. This equation can be written in matrix form:
\begin{equation}
    \label{eq:sys_matrix}
    \partial_t\delta\comp=H\delta\comp
\end{equation}
where \(\delta \comp=
\begin{pmatrix}
\delta \comp_1\\\vdots\\\delta \comp_{N_z}
\end{pmatrix}\) and \(H\) is the symmetric circulant matrix:
\[H=
\begin{pmatrix}
A & B & C & 0 & \cdots & 0 & C & B\\
B &   & \ddots & \ddots & \ddots &   & 0 & C \\
C & \ddots & \ddots &   & C & 0 &   & 0 \\
0 & \ddots &   & A & B &   & \ddots & \vdots \\
\vdots & \ddots & C & B &  &   & \ddots & 0 \\
0 &   & 0 &   &   & \ddots & \ddots & C \\
C & 0 &   & \ddots & \ddots &\ddots &   & B \\
B & C & 0 & \cdots & 0 & C & B &  A
\end{pmatrix}
\]
where
\begin{align}
    A=&\nabla\cdot M\nabla\left(-\kappa\nabla^2+g''+2\partial_{20}\fNN\right.\notag\\
    &\left.+2\partial_{200} \fNNN+\partial_{020}\fNNN\right)
\end{align}
\begin{align}
    B=\nabla\cdot M\nabla\left(\partial_{11}\fNN+2\partial_{110}\fNNN\right)
\end{align}
\begin{align}
    C=\nabla\cdot M\nabla\partial_{101} \fNNN.
\end{align}
The circulant structure is due to the invariance of the system with respect to discrete translation along $z$ by one layer.

The system (\ref{eq:sys_matrix}) is a partial differential equation with a time derivative on the left-hand side and space derivatives on the right-hand side. Separation of variables then yields that the solution is exponential in time. Furthermore, the translation invariance in space implies that the spatial solution can be expanded in terms of plane waves. Therefore, the general solution is of the form
\begin{equation}
    \delta \comp(x,t)=\sum_{m=0}^{N_z-1}\sum_{n=-\infty}^{+\infty} \frac 12 \left(a_{mn} e^{ik_nx}e^{\omega_m(k_n)t}v_m \; +c.c.\right)
    \label{eq:decompo}
\end{equation}
where \(v_m\) are the eigenvectors of the matrix \(H\), \(k_n=2\pi n/L\) is the wavenumber along $x$ (the discrete values are due to the finite system size), \(\omega_m(k_n)\) is the growth rate associated to the eigenvector $v_m$ and wavenumber $k_n$, and $a_{mn}$ are the amplitudes of each mode.
In the following, we will assume that the system size $L$ is large enough for the $k_n$ to form a quasi-continuum, and drop the index $n$.
With these notations, the growth rates are directly related to the eigenvalues \(\lambda_m\) of the matrix \(H\):
\begin{align}
    \omega_m(k)=\lambda_m(k)
\end{align}
The number of different eigenvectors equals the number of layers.
The eigenvectors of a circulant matrix are the discrete plane waves along \(z\):
\begin{align}
    v_m=\dfrac{1}{\sqrt{N_z}}\begin{pmatrix} 1 \\\exp\left(\dfrac{2i\pi m}{N_z}\right)\\\vdots\\\exp\left(\dfrac{2i\pi m(N_z-1)}{N_z}\right)\end{pmatrix}.
    \label{eq:eigenvectors}
\end{align}

Introducing the solution (\ref{eq:decompo}) in the system (\ref{eq:sys_matrix}), we can express the coefficients \(A\), \(B\) and \(C\) of the matrix as function of the wavenumber $k$: 
\begin{align}
    A(k)&=-Mk^2\left(\kappa k^2+g''+2\partial_{20}\fNN\right.\notag\\
    &\left.+2\partial_{200} \fNNN+\partial_{020}\fNNN\right)
\end{align}
\begin{align}
    B(k)=-Mk^2\left(\partial_{11}\fNN+2\partial_{110}\fNNN\right)
\end{align}
\begin{align}
    C(k)=-Mk^2\partial_{101} \fNNN
\end{align}
As the matrix \(H\) is a symmetric circulant matrix, the expression for the eigenvalues is easy to find and reads
\begin{align}
    \lambda_m(k)=A(k)+2B(k)\cos{\dfrac{2\pi m}{N_z}}+2C(k)\cos{\dfrac{4\pi m}{N_z}}.
\end{align}

Although the number of layers in a graphite particle may be very large, we can  consider a much smaller number of layers in the analysis depending on the number of stages we want to study.
In the following, to detail the expressions of the different growth rates, we consider \(N_z=6\) layers which is compatible with the appearance of both stages 2 and 3. It is useful here to detail the relations between eigenvectors and stages, and to label the growth rates by the stage number rather than the index of the eigenvalue.

The eigenvector associated to the mode $m=0$ is uniform (all elements are $1$). 
The corresponding growth rate therefore describes the evolution of the local average concentration. 
Since this has the same symmetry as stage 1 (all layers behave the same way) we label the corresponding growth rate with index $1$.  
\begin{align}
    \omega_1(k)&=A+2B+2C\notag\\
    & =-Mk^2\bigg\{\kappa k^2+N_V\bigg[\dfrac{k_B T}{\bar{c}(1-\bar{c})}\notag\\
    & -2\Omega_a+2\Omega_b-2\Omega_c(3\bar{c}-1)\bigg]\bigg\}.
    \label{eq:gr_rat_avg}
\end{align}
The modes $m=1$ and $m=5$ are degenerate due to the up-down symmetry. The corresponding eigenvectors correspond to discrete plane waves, in the $z$ direction, of wavelength $6 L_z$, propagating in the positive and negative $z$ direction, respectively. Therefore, these modes would correspond to a stage 6, which does not appear in the phase diagram. The corresponding growth rate is
\begin{align}
    \omega_6(k)&=A+B-C\notag\\
    & =-Mk^2\bigg\{\kappa k^2+N_V\bigg[\dfrac{k_B T}{\bar{c}(1-\bar{c})}\notag\\
    & -2\Omega_a+\Omega_b-\Omega_c(\bar{c}+1)\bigg]\bigg\}
    \label{eq:gr_rat_6}
\end{align}
Similarly, the modes $m=2$ and $m=4$ are also degenerate, and the corresponding eigenvectors are discrete plane waves of wavelength 3; these modes are therefore associated with stage 3, and their growth rate is
\begin{align}
    \omega_3(k)&=A-B-C\notag\\
    & =-Mk^2\bigg\{\kappa k^2+N_V\bigg[\dfrac{k_B T}{\bar{c}(1-\bar{c})}\notag\\
    & -2\Omega_a-\Omega_b+\Omega_c(3\bar{c}-1)\bigg]\bigg\}
    \label{eq:gr_rat_3}
\end{align}
Finally, the last mode, $m=3$, has an eigenvector proportional to $(-1)^j$ and corresponds to stage 2, with a growth rate of
\begin{align}
    \omega_2(k)&=A-2B+2C\notag\\
    & =-Mk^2\bigg\{\kappa k^2+N_V\bigg[\dfrac{k_B T}{\bar{c}(1-\bar{c})}\notag\\
    & -2\Omega_a-2\Omega_b+2\Omega_c(\bar{c}+1)\bigg]\bigg\}
    \label{eq:gr_rat_2}
\end{align}

\begin{figure}
    \centering
    \includegraphics[width=\linewidth]{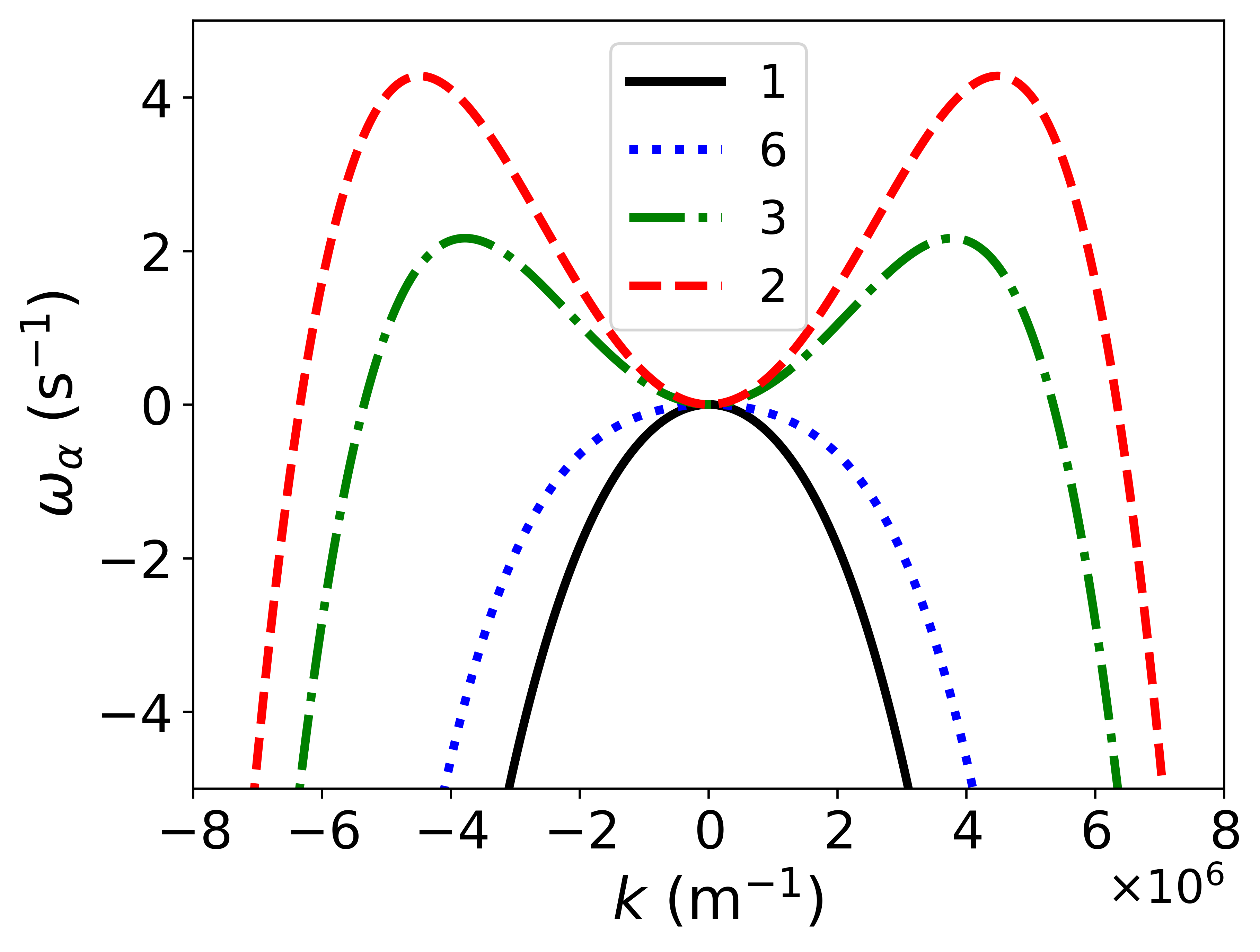}
    \caption{Growth rates for the different stages for an average composition of \(\bar{c}=0.3\), a temperature of \(T=298~K\) and the thermodynamic parameters of Table~\ref{tab:phase_diagram}.}
    \label{fig:GrowthRate}
\end{figure}

An example for a stability spectrum (the growth rates of the different modes as a function of the wavenumber $k$) is displayed in Fig.~\ref{fig:GrowthRate} for the parameters from Table~\ref{tab:phase_diagram}, and for an average composition \(\bar{c}=0.3\), at \(T=298~K\).
A mode is unstable (and a phase decomposition is observed) when its growth rate is positive. For example, for stage \(2\), the system is unstable for wavenumbers between \(-6.34\times 10^6~m^{-1}\) and \(6.34\times 10^6~m^{-1}\). On the contrary, we can see that stage \(6\) is stable for all wavenumbers for this temperature and average composition.
For this mean filling fraction of \(\bar{c}=0.3\), the growth rate of stage \(2\) is higher than the one of stage \(3\).
This is remarkable because, according to the phase diagram, stage \(3\) should be present at equilibrium. The theoretical analysis thus predicts that stage \(2\) should emerge during the first few instants of a spinodal decomposition right after a quench, even though the final state must be stage $3$.

\subsection{Characteristic wave numbers and growth rates}
For each mode (stage), there are two noteworthy wavenumbers: the one where the growth rate reaches its maximum value, \(k_{\rm max}\); and the marginal one for which the growth rate is zero, \(k_0\).
To compute these numbers, the growth rates are rewritten in the general form
\begin{align}
    \omega_\alpha(k)=-Mk^2[\kappa k^2-\Gamma_\alpha(\bar{c})]
\end{align}
for \(\alpha=1\), \(2\), \(3\) or \(6\), where \(\Gamma_\alpha\)
\begin{eqnarray}
\Gamma_1 & = & N_V\bigg[-\dfrac{k_B T}{\bar{c}(1-\bar{c})}+2\Omega_a-2\Omega_b+2\Omega_c(3\bar{c}-1)\bigg]\notag\\
\Gamma_2 & = & N_V\bigg[-\dfrac{k_B T}{\bar{c}(1-\bar{c})}+2\Omega_a+2\Omega_b-2\Omega_c(\bar{c}+1)\bigg]\notag\\
\Gamma_3 & = & N_V\bigg[-\dfrac{k_B T}{\bar{c}(1-\bar{c})}+2\Omega_a+\Omega_b-\Omega_c(3\bar{c}-1)\bigg]\notag\\
\Gamma_6 & = & N_V\bigg[-\dfrac{k_B T}{\bar{c}(1-\bar{c})}+2\Omega_a-\Omega_b+\Omega_c(\bar{c}+1)\bigg]\notag
\end{eqnarray}
is the part that originates from interactions and entropy in Eqs.~(\ref{eq:gr_rat_avg}) to (\ref{eq:gr_rat_3}).
Computing the \(k_0\) and \(k_{\rm max}\) values only makes sense for unstable states, when \(\Gamma_\alpha(\bar{c})>0\).
For \(k_0\), one needs to find where \(\omega_\alpha(k)=0\) with \(k\neq 0\),
\begin{equation}
    k_0^{(\alpha)}=\sqrt{\dfrac{\Gamma_\alpha}{\kappa}}
    \label{eq:k0}
\end{equation}
Figure~\ref{fig:GrowthRate} shows that \(k_{\rm max}\) is found when the first derivative of \(\omega_\alpha(k)=0\) and \(k\neq 0\),
\begin{equation}
    k_{\max}^{(\alpha)}=\sqrt{\dfrac{\Gamma_\alpha}{2\kappa}}. 
    \label{eq:kmax}
\end{equation}
The corresponding maximal growth rate is
\begin{equation}
\omega_{\max}^{(\alpha)} = M\frac{\Gamma_\alpha^2}{4\kappa}.
\label{eq:omegamax}
\end{equation}
The mode corresponding to this maximal growth rate will dominate the emerging pattern; therefore, the inverse of its growth rate gives a characteristic time scale for phase separation, which we will denote by $\tau_{\rm dec}$, and which depends on the initial composition,
\begin{equation}
    \tau_{\rm dec} = \frac{4\kappa}{M\Gamma_\alpha(\bar c)^2}
\label{eq:decomposition_time}
\end{equation}

The growth rates of the different stages depend on the interaction energies (\(\Omega_a\), \(\Omega_b\) and \(\Omega_c\)) through the coefficients $\Gamma_\alpha$. The condition $\Gamma_\alpha(\bar c)>0$ determines the composition range in which mode $\alpha$ is unstable. Since the growth rate increases with $\Gamma_\alpha$, the mode with the highest $\Gamma$ grows the fastest and has the widest unstable composition range. Of particular interest is the competition between stages $2$ and $3$. The condition $\Gamma_2>\Gamma_3$ is equivalent to
\begin{equation}
    \dfrac{\Omega_c}{\Omega_b}<\frac{1}{3-\bar{c}}.
\label{eq:stagesorder}
\end{equation}
For the parameters of Table~\ref{tab:phase_diagram}, this is the case for all values of $\bar c$, such that stage \(2\) always grows the fastest at short times.
It can also be shown that for these parameters, the onset of spinodal decomposition, obtained by solving $\Gamma_{\alpha}(\bar{c}) = 0$, occurs for lower composition for stage 2 than for stage 3. 
The competition of growth rates between stages 1 and 2 deserves also some attention. The condition $\Gamma_2>\Gamma_1$ is equivalent to 
\begin{equation}
    \dfrac{\Omega_c}{\Omega_b}<\frac{1}{2\bar{c}}.
\label{eq:stagesorder1}
\end{equation}
If we consider all the possible filling fraction, this condition reduces to $\Omega_c< 2 \Omega_b$, which is fulfilled for the parameters of Table~\ref{tab:phase_diagram}. This means that stage 2 will always appear before any potential decomposition between stage 1 and stage 1'.

As will be shown below (Section~\ref{sec:discussion}), these conclusions remain valid for other ratios of the inter-layer free-energies $\Omega_c$ and $\Omega_b$, as long as the ratio leads to the right staging sequences.

\section{Numerical simulations\label{sec:num}}
\subsection{Numerical method}
To study the kinetics of spinodal decomposition beyond the early times during which the linearization is valid, numerical simulations are needed. The free-energy model of section~\ref{sec:free-energy} is introduced in the multi-layer formulation of Eq.~(\ref{eq:Cahn_Hilliard}). As in the theoretical analysis, a system of 6 layers makes it possible to capture both stage \(2\) and stage \(3\).
Periodic boundary conditions are used in the $z$ direction. The system is supposed to be invariant along the $y$ direction. The particle size is fixed to 
\(25\;\rm{\mu m}\), which corresponds to a typical size of graphite particles in standard electrodes.
The value of $D$ is taken from Ref.~\cite{smith_intercalation_2017}, in coherence with the dependence of the mobility on the local lithium filling fraction (see Eq.~\ref{eq:mobilite}). 

An important issue is the choice of the gradient penalty term $\kappa$. In a mean-field approximation, its value is related to the strength and spatial range of attrative in-plane interaction between lithium atoms \cite{LangerBegRohu}; an order-of magnitude estimation of $10^{-10}~\rm{J} \; \rm{m}^{-1}$ has been given in Ref.~\cite{Han2004}. Since the gradient enegy penalizes spatial variations of the concentration, it directly influences the thickness of the diffuse interfaces between concentrated and dilute domains in the same layer. This thickness, which can be estimated as \(\lambda\approx \sqrt{\kappa/(\Omega_a N_V)}\), is of the order of a few nanometers for the quoted value of $\kappa$. The thickness of the interfaces yields a practical constraint, since the spacing of the numerical grid must be smaller than the interface thickness in order to properly resolve the interfaces. For this reason, a particle of size \(25\;\mu m\) cannot be simulated with this value of $\kappa$ within reasonable simulation times. Therefore, in the present work, a larger value is used, $\kappa = 3 \times 10^{-6}~\rm{J}\; \rm{m}^{-1}$. Some consequences of this choice will be discussed below.

A finite-volume method is used to discretize the equations in the $x$-direction.
The grid spacing is chosen to ensure that it is small enough to capture the interfaces between lithium-poor and lithium-rich domains. Given the stiffness of the problem and to keep the computational time reasonable the time integration is carried out using  an implicit method. The nonlinear system of equations arising from the implicitation is solved using a Newton method combined with a GMRES solver for the resolution of the linear problem~\cite{Brown1990,Choquet1996}. The algorithm adapted from~\cite{Brown1990,Choquet1996} has been implemented in an in-house code, originally developed for the study of diffuse interface models in fluid mechanics~\cite{Jamet2005}.

Spinodal decomposition occurs when a homogeneous phase is linearly unstable and spontaneously separates into two phases.
The simulation of this phenomenon is possible by imposing a zero lithium flux at both $x=0$ and $x=L$ and by initializing the simulation with an homogeneous filling fraction $\bar c$ in all the layers.
Since the decomposition cannot happen if the composition is perfectly homogeneous, a small perturbation $\delta \tilde c_j(x)$ is added to the average initial composition. 
In the generic case a randomly uniformly distributed perturbation of up to $\pm 5 \%$ was added to each grid point.  For the comparison between simulation and theory the initial conditions are given below.


\begin{figure}[ht!]
    \centering
    \includegraphics[width=\linewidth]{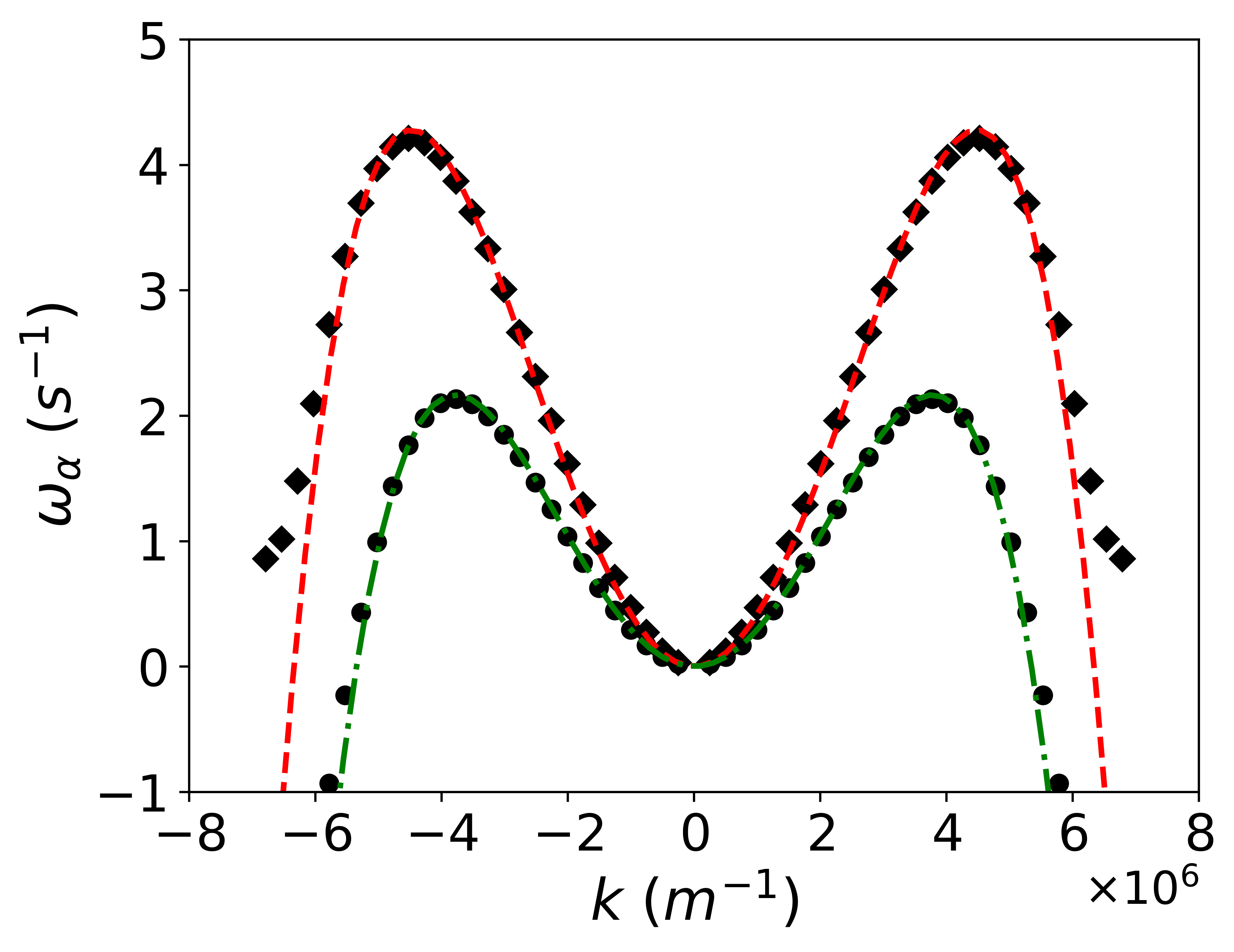}
    \caption{Numerical (symbols) and analytical (lines) growth rates for stage 2 (diamonds and dashed red line) and stage 3 (circles and dash-dotted green line) for an initial composition \(\bar{c}=0.3\), a temperature $T=298K$ and the parameters of TABLE~\ref{tab:phase_diagram}.}
    \label{fig:Numerical_Theoretical}
\end{figure}

\subsection{Validation: linear regime}
A comparison between the theoretical growth rates predicted by the linear stability analysis and the simulated ones has been performed. Since the linear stability analysis is valid only for small \(\delta\comp_j\), the comparison is performed at short times after the quench. Two successive Fourier transforms are applied to the composition fields obtained from the simulations. The first is a discrete Fourier transform in the $z$ direction,
\begin{equation}
    \hat{c}_m(x,t)=\dfrac{1}{N_z}\sum_{j=0}^{N_z-1}\comp_j(x,t)e^{iq_mj}
    \label{eq:DiscreteFourierTransform}
\end{equation}
where \(q_m=2\pi m/N_z\) is the discrete wave number in the $z$-direction.
This post-treatment allows us to distinguish the stages at all times and depths in the graphite particle. 
Furthermore, we perform a continuous Fourier transform along the $x$ direction,
\begin{equation}
    \bar{c}_{mn}(t)=\dfrac{1}{L}\int_{0}^{L}\hat{c}_m(x,t)e^{ik_nx}dx
    \label{eq:ContinuousFourierTransform}
\end{equation}
where \(k_n=2\pi n/L\) is the wavenumber along \(x\), which takes discrete values due to the finite system size. The combination of the two transforms yields the Fourier coefficient of mode number $m$ with wavenumber $k_n$ along $x$,
\begin{equation}
    \bar{c}_{mn}(t)=\dfrac{1}{N_z L}\int_{0}^{L}\left(\sum_{j=0}^{N_z-1}\comp_j(x,t)e^{iq_mj}\right)e^{ik_nx}dx
    \label{eq:SuccessiveFourierTransform}
\end{equation}
In the numerical implementation, Simpson's rule has been used to compute the integral.
In the regime immediately after the quench, the logarithm of the amplitude $|\bar{c}_{mn}(t)|$ varies linearly with time, and a fit of the slope provides the growth rate. 

Using the linear superposition principle, it should be theoretically possible to extract the growth rates of all the modes from a single simulation with a random initial fluctuation. However, the precision is deteriorated by roundoff errors, and therefore simulations were performed for each individual mode and each wavelength.   
All the simulations have an average initial composition of \(\bar{c}=0.3\), as in the results of Fig.~\ref{fig:GrowthRate}.
The perturbations corresponds to a prescribed mode and a prescribed plane wave. 
For stage two, the initial compositions are:
\begin{align}
\comp_j(x,0)=\bar{c} + \dfrac{\cos(k_nx+j\pi)}{1000}\quad\text{with}~n=[0,\cdots,27],
\label{eq:IC-II}    
\end{align}
while for stage three:
\begin{align}
\comp_j(x,0)=\bar{c} + \dfrac{\cos(k_nx+2j\pi/3)}{1000} \notag \\ \text{with}~n=[-23,\cdots,23].
\label{eq:IC-III}    
\end{align}
These expressions correspond, respectively, to the eigenvectors $m=3$ and $m=2$ of Eqs.~(\ref{eq:decompo}) and (\ref{eq:eigenvectors}), with amplitudes $a_{3n}=a_{2n}=1/1000$.
The number of simulated modes was determined as a function of system size and the theoretically calculated marginal wavenumber such as to cover all the relevant range of the stability spectrum.
The amplitude of the fluctuations ($1/1000$) was small enough to ensure a sufficiently long duration of the linear regime for obtaining good fits of the growth rates.
A total number of 75 simulations have been performed to analyse the growth rate of each individual mode and each wavelength.
Figure~\ref{fig:Numerical_Theoretical} shows the comparison of the growth rates obtained from the numerical simulations and from the linear stability analysis. The excellent agreement between theory and simulation confirms the validity of the linear stability analysis as well as the implementation of the numerical model.

\begin{figure*}[ht!]
    \centering
    \includegraphics[width=\textwidth]{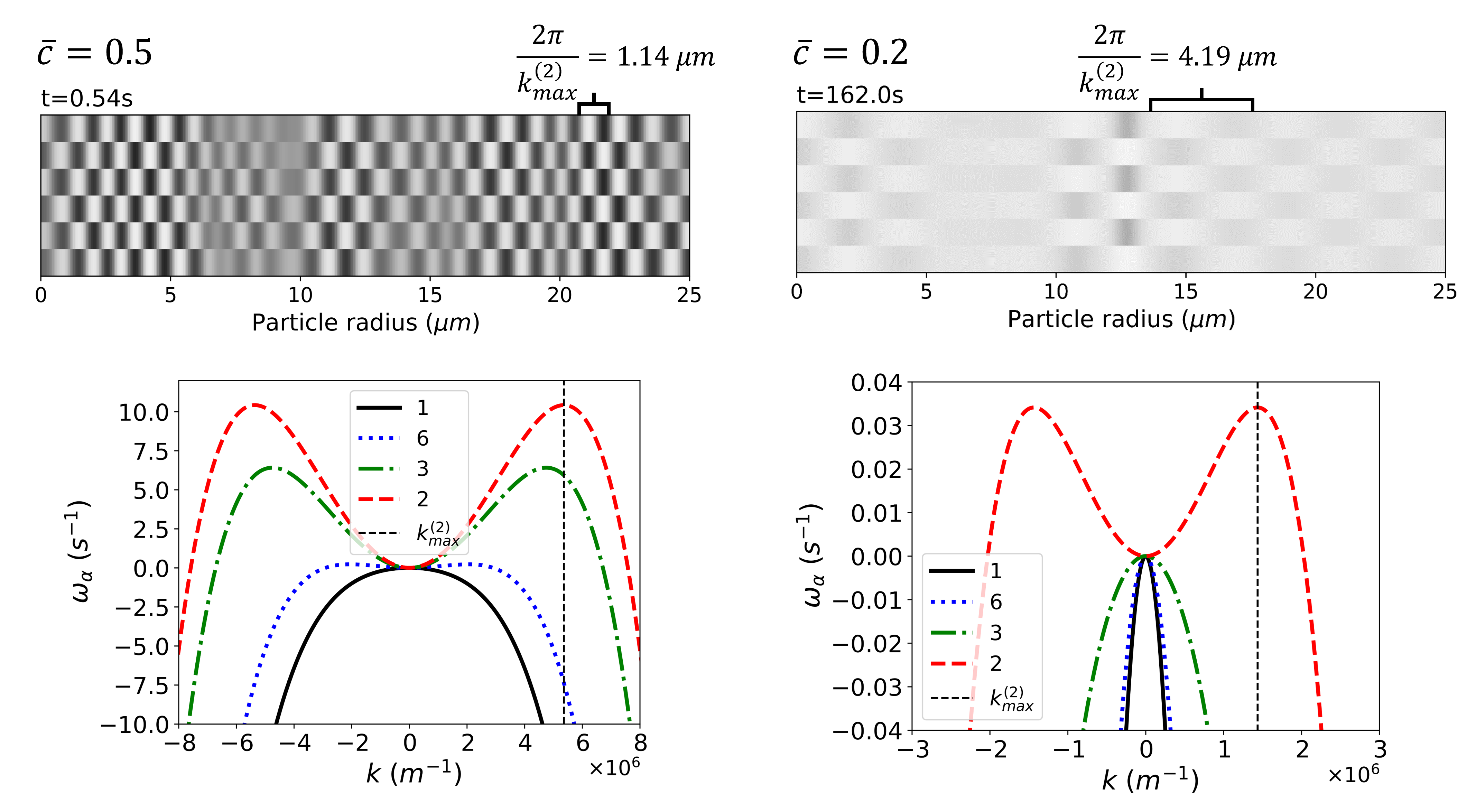}
    \caption{Top: Snapshots of spinodal decompositions for two different initial mean filling fraction : \(\bar{c}=0.5\) (left) and \(\bar{c}=0.2\) (right). The snapshots are taken slighty after the end of the linear regime to achieve a sufficient contrast between dilute and dense domains. Bottom: Corresponding linear stability analysis growth rate curves with an emphasis on the wave number for which the stage 2 is at its maximum, \(k_{max}^{(2)}\) (black dashed lines).}
    \label{fig:domain_size}
\end{figure*}
All the plane waves with \(k < k_0\) will grow exponentially in time, the fastest one being the closest to \(k_{\rm max}\). Therefore, at the end of the linear stage the size of the lithium islands should be of the order of \(2\pi/k_{\rm max}\). 
This prediction can be verified by performing simulations of spinodal decompositions with random initial perturbation for two different initial mean filling fractions, \(\bar{c}=0.5\) and \(\bar{c}=0.2\).
The stability spectra are presented in Fig.\ref{fig:domain_size} together with snapshots of the simulations at early times. For \(\bar{c}=0.5\), \(k_{\rm max}\) is expected to be larger than for \(\bar{c}=0.2\), leading to smaller islands. This is consistent with the snapshot pictures, where the wavelength of the fastest growing mode, $2\pi/k_{\rm max}$, is displayed for comparison. The agreement is very satisfactory.

The linear stability analysis also predicts the growth rate of the perturbations.
At \(\bar{c}=0.5\), the growth rate of the leading mode \(\omega_2(k_{\rm max}) \) is of the order \(10.5~s^{-1}\), corresponding to a characteristic decomposition time $\tau_{\rm dec} \approx 0.1s$, in coherence with the formation of well-developed domains in less than half a second, as can be seen Fig.~\ref{fig:domain_size} on the left.
At lower filling fraction, \(\bar{c}=0.2\), the growth rate of the leading mode is much lower, \(\omega_2(k_{\rm max})\approx 0.034~s^{-1}\) corresponding to a characteristic decomposition time $\tau_{\rm dec} \approx 30s$. Even after 162 seconds, one can barely identify the domains in the snapshot picture of Fig.~\ref{fig:domain_size} on the right. While for \(\bar{c}=0.3\), stage 6 is stable for any wave number  (Fig.~\ref{fig:GrowthRate}), we can notice that it is slightly unstable for \(\bar{c}=0.5\) as can be noticed on Fig~\ref{fig:domain_size} (left). However, since it is not the dominant unstable mode, it will not easily be observed.  

\subsection{Coarsening: the non-linear regime} 

\begin{figure*}[ht!]
    \centering
    \includegraphics[width=0.85\textwidth]{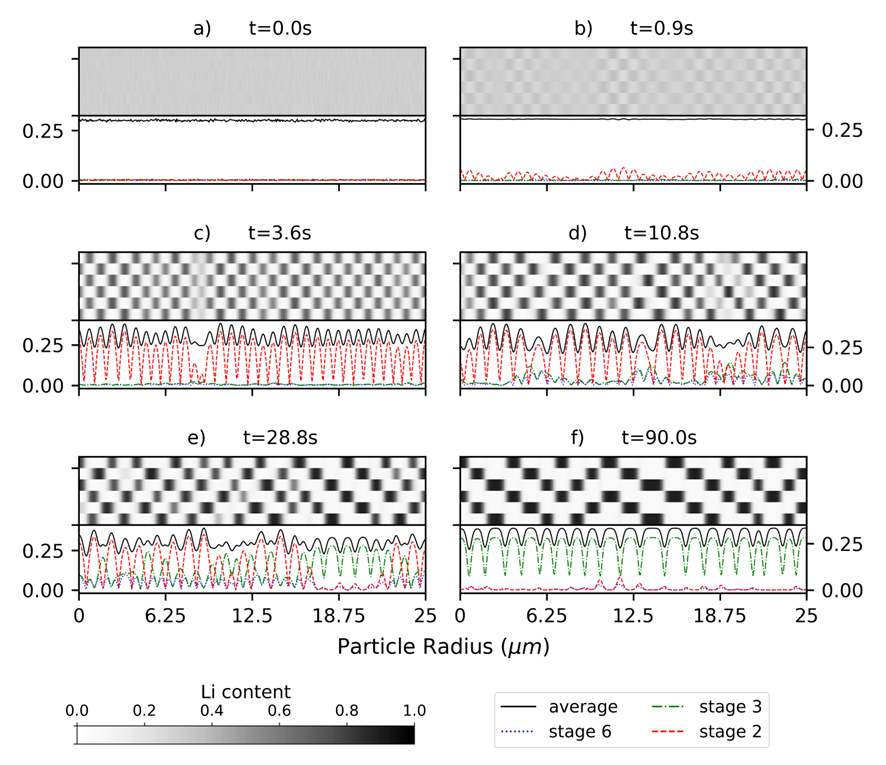}
    \caption{Snapshots of a spinodal decomposition at different times (see also movie in the Supplemental Material \protect\cite{MovieFig6}). The initial average composition is \(\bar{c}=0.3\) and the system has \(N_z=6\) layers. For each subfigure, top: lithium content in each layer, bottom: coefficients of the Fourier transform \(|\hat{c}_m(x,t)|\) (see Eq.~\ref{eq:DiscreteFourierTransform}).}
    \label{fig:decomposition}
    
\end{figure*}
Figure \ref{fig:decomposition} presents snapshots at different times of the spinodal decomposition starting from an initial filling fraction of \(\bar{c}=0.3\) with a random initial fluctuation. 
At the top part of each subfigure, the lithium content in all of the 6 layers and for the whole particle depth can be seen in the form of a grey-scale heat-map.
The bottom part displays the result of the discrete Fourier transform (Eq.~(\ref{eq:DiscreteFourierTransform})): the amplitudes of the Fourier coefficients can be used to identify quantitatively the stages present at each position within the particle and thus to follow the evolution of the system beyond the linear regime. 

The particle starts from a homogeneous state with small fluctuations (a). Very quickly, all the layers undergo spinodal decomposition. As predicted by the linear stability analysis, stage $2$ grows the fastest, and a corresponding checkerboard pattern starts to appear in the heat map, whereas the average composition still remains homogeneous (b). The amplitude of this pattern increases, until domains of high (close to $\tilde c=1$) and low (close to $\tilde c=0$) composition clearly emerge and form vertical columns (c). The average composition is now higher in the center of the columns than in the frontier between two columns. This is a clear indication of a non-linear interaction between different modes, and thus the 
linear stability analysis and Eq.~(\ref{eq:CH+Free_NRJ_lin}) do not apply anymore.

Since the simulation was started from a random perturbation, the size of the islands (columns) is not everywhere the same, and the pattern is slightly disordered. This subsequently leads to coarsening events: some domains expand in size, while others disappear (d). During this coarsening, stage 3 progressively appears, even though the global average filling fraction stays constant (e). This is visible on the greyscale map, but also on the plot of the local Fourier coefficients. One particular mechanism by which stage 3 appears in regions previously occupied by stage 2 is the following: the disappearance of one concentrated domain in a single layer is followed by a sequence of elimination events in the other layers which makes an entire ``diagonal row'' of concentrated domains disappear. Subsequently, the domains located on both sides of the eliminated row grow closer to each other, until they locally reach a stage 3 pattern. Several such events can be seen in the movie included in the Supplemental Material \cite{MovieFig6}, for example in the center of the system at $t\approx 100$s. At the end of the simulation, the entire system is in stage 3 (f).

The column structure is nicely represented by the Fourier coefficients $\hat{c}_m(x,t)$ defined in Eq.~(\ref{eq:DiscreteFourierTransform}). They are complex numbers, with an amplitude and a phase; only the amplitude is plotted in Fig.~\ref{fig:decomposition}. In stage 2, two neighboring columns have the same amplitude, but a phase difference of $\pi$. In stage 3, two neighboring columns can have a phase difference of $\pm 2\pi/3$, which correspond to the ``upward and downward staircases'' visible in Fig.~\ref{fig:decomposition}; these two different states reflect the two degenerate linear modes that contribute to stage 3. In the Fourier representation, the boundary between two columns is thus a ``phase jump''. Since, in each layer, the composition variation in space is controlled by the gradient energy coefficient, the thickness of such a domain wall should be close to the one of an interface between dilute and concentrated domains in a single layer.

\begin{figure}[h!]
    \centering
    \includegraphics[width=\linewidth]{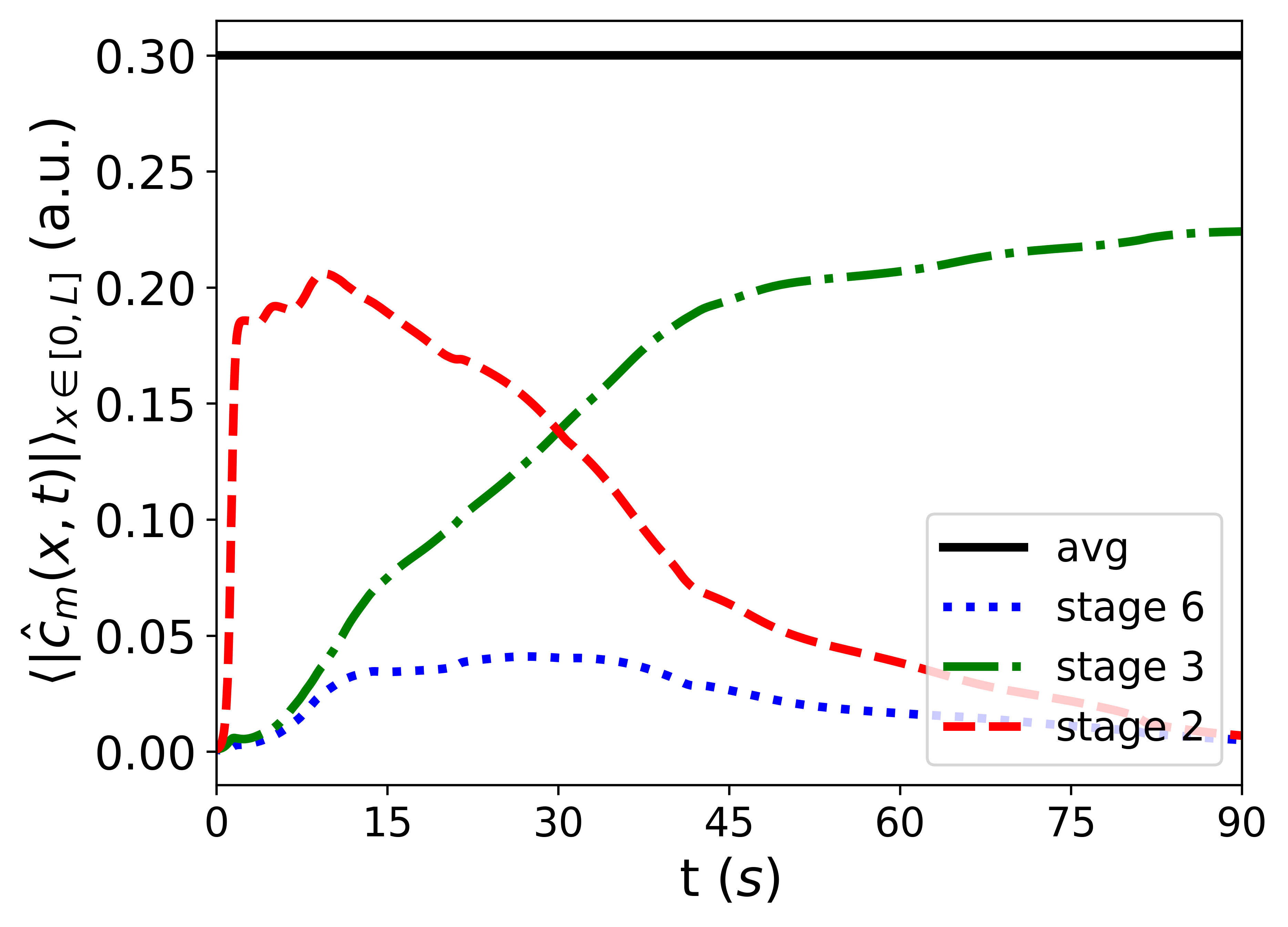}
    \caption{Time evolution of the global stage fractions \(\langle|\hat{c}_m(x,t)|\rangle_{x\in[0,L]}\) during the spinodal decomposition at \(T=298K\), with an initial composition of \(\bar{c}=0.3\).}
    \label{fig:modes}
\end{figure}

To follow quantitatively the global presence of the different stages during the coarseing, the amplitude of the Fourier coefficients \(\hat{c}_m(x,t)\) can be averaged over the length of the particle from \(x=0\) to \(x=L\).
The time evolution of these averages, noted \(\langle|\hat{c}_m(x,t)|\rangle_{x\in[0,L]}\), are presented in Fig. \ref{fig:modes}.
This figure clearly shows that stage \(2\) grows the fastest at early times in accordance with the linear stability analysis. After a few seconds, stage \(3\) appears and increases to finally dominate after 45s, as expected from the result of the phase diagram for an initial mean filling fraction of \(\bar{c}=0.3\). 

Simulations with the multi-layer Cahn-Hilliard framework are therefore a powerful tool to analyze the dynamics of stage
formation and evolution, and to bridge the gap between the early-time dynamics predicted by the linear stability analysis and the final equilibrium state predicted by the phase diagram. 

\section{Discussion\label{sec:discussion}}

\begin{figure*}[ht!]
    \centering
    \includegraphics[width=0.9\textwidth]{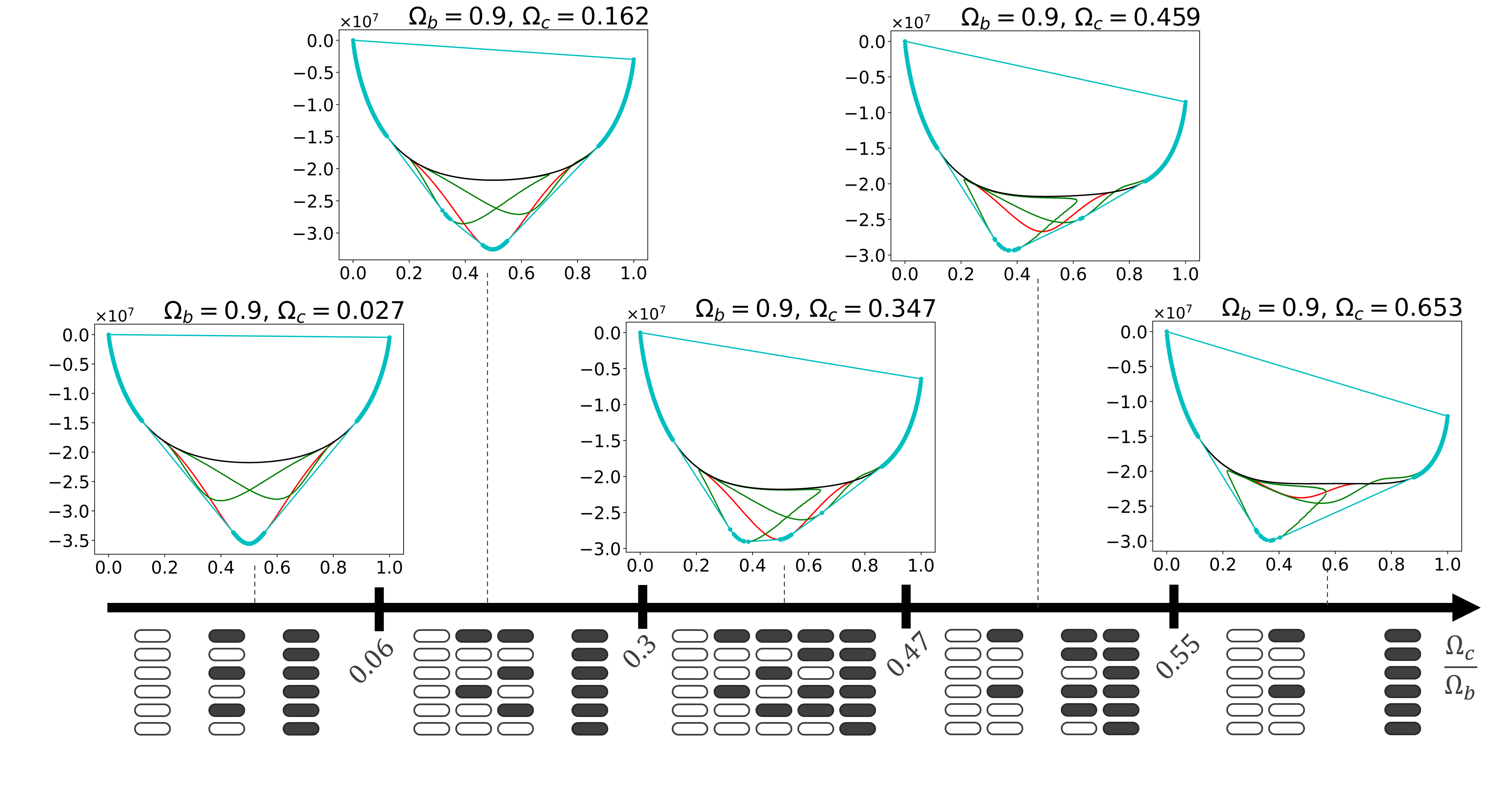}
    \caption{Sequence of stages occurring with increasing filling fraction, for various ratio $\Omega_c/\Omega_b$. The ratio $\Omega_c/\Omega_b$ increases from left to right. The values of $\Omega_b$ and $\Omega_c$ are non-dimensionalized using the prefactor $k_BT$ at $298K$. Above the arrow: free-energy curves of all the stages (black: stage 1 or dilute, red: stage 2, green: stage 3 and 3/2) together with the convex envelope (cyan). Below the arrow: resulting scheme of the  staging sequence. Ratio $\Omega_c/\Omega_b$ below 0.06: no stage 3; between 0.06 and 0.3: right sequence, between 0.3 and 0.47: occurrence of stage 3/2; between 0.47 and 0.55: no stage 2 and occurrence of stage 3/2; above 0.55 no stage 2.}
    \label{fig:omega_c_omega_b}
\end{figure*}

The most noteworthy result of the preceding analytical and numerical studies is that, for the interaction energy parameters listed in Table~\ref{tab:phase_diagram}, stage 2 always grows faster than stage 3, even in the composition range where stage 3 is stable at thermodynamic equilibrium. As already stated in Eq.~(\Ref{eq:stagesorder}), stage 3 would be favored if the ratio $\Omega_c/\Omega_b$ was large enough. 
However, in Ref.~\cite{chandesris_thermodynamics_2019} it was found that in order to obtain the right sequence of stages for increasing filling fraction (and, in particular, to avoid the occurrence of the non-physical stage 3/2), $\Omega_c/\Omega_b$ has to lie within a certain range. 
We have not found a closed-form analytic expression for this latter condition, but we have numerically investigated the phase diagram for a large range of $\Omega_b$ and $\Omega_c$.
As an example, we show in Fig~\ref{fig:omega_c_omega_b} the free-energy curves for all the stages together with their convex envelopes and the corresponding stage sequence obtained for a fixed ratio of $\Omega_a/\Omega_b = 2.78$, a fixed temperature of 298K, and various $\Omega_c/\Omega_b$ ratio.
Stage 3/2 appears before the condition $\omega_3 > \omega_2$ is satisfied.
We have not found any set of values for which both the equilibrium sequence of stages was correct and stage 3 grew the fastest after a quench. Therefore, we must conclude that, in the framework of a free-energy model of the form of Eqs.~(\ref{eq:NRJ_inter_NN}) and (\ref{eq:NRJ_inter_NNN}), the fast growth of stage 2 in the spinodal decomposition of a homogeneously filled sample is a general phenomenon.
This methodology could easily be extended to other intra- and interlayer interactions energies~\cite{smith_intercalation_2017, Safran1980, hawrylak_kinetic_1984} to determine if this finding holds for an even wider class of free-energy models.

Let us now discuss the consequences of our findings for the situation, more relevant for comparisons to experiments, in which an initially empty electrode is progressively charged. The active material initially is in the dilute stage 1. Lithium atoms are inserted into the material at the interface with the electrolyte, and the distribution of atoms inside the material proceeds by diffusion. There are two distinct characteristic time scales~\cite{Nadkarni2018,Fraggedakis2020}: the diffusion time $\tau_D\sim L^2/D$, and the processing time scale (charging time) $\tau_I\sim Q/iA$, where $i$ is the externally imposed charging current density, $Q$ is the total capacity of the active material, and $A$ is the surface area over which the current is applied. For slow charging, such that $\tau_I\gg\tau_D$, diffusion is fast enough to distribute the lithium atoms through the entire particle, and the concentration remains almost homogeneous. The global concentration thus slowly increases until it reaches the limit where spinodal decomposition can occur. As already discussed above, the first mode that becomes unstable is stage 2. Therefore, the system should undergo a sequence of states that is comparable to our simulations of Fig.~\ref{fig:decomposition}. It should be noted that the decomposition time given by Eq.~(\ref{eq:decomposition_time}) depends on the concentration and is infinite at the spinodal. Therefore, the decomposition will become visible at a concentration that is larger than the critical concentration for instability, and that depends on the charging rate. Indeed, the instability will occur approximately when the concentration is such that $\tau_{\rm dec}(\bar c)=\tau_I$. 

In the opposite limit, where $\tau_I<\tau_D$, inhomogeneities in concentration develop inside the active material. Since we start from a dilute system in which the diffusion is well described by Fick's law, the vicinity of the electrolyte will always be the point with the highest concentration, and will therefore reach the threshold for instability first. Since the composition threshold for the onset of spinodal decomposition is the lowest for stage 2, and its growth rate always remains the highest for compositions beyond the threshold, the present theory predicts that stage 2 forms at the surface of the electrode. This was indeed observed in the preliminary simulations of the charging process in Ref.~\cite{chandesris_thermodynamics_2019}. Whether this surface layer of stage 2 persists or disappears in the subsequent evolution is an important question, because the occurrence of staging at the surface will modify the effective resistance for the entry of new atoms into the material. To address this issue, interfacial intercalation kinetics with its potential nonlinearities must also be taken into account \cite{Bazant2017,Bazant2023}.
This issue will be adressed numerically in a forthcoming publication through detailed simulations of the charge-discharge dynamics. 

Furthermore, these results imply that at high charging rates, stage 3 should disappear completely. This has not been reported experimentally so far, but could be confirmed using high speed synchrotron X-ray diffraction~\cite{Yao2019Gr,Finegan2020,Schmitt2021}. It would require to specifically follow the reflections characteristics of the stage 3 ordering, i.e. the examination of reflections on planes separated by about 3 times the graphite interlayer distance d. This is possible at low charging rates ~\cite{Tardif2021a}, but remains challenging at higher C-rate, given their low intensity.

As already mentioned, we have used for our simulations a value of the gradient energy coefficient $\kappa$ that is much larger than the one expected from a mean-field approximation. This is an issue because $\kappa$ appears in the expression for the growth rates and for the characteristic wavelengths, Eqs.~(\ref{eq:kmax}) through (\ref{eq:k0}). The typical initial size of the domains is $2\pi/k_{\rm max}$; therefore, a measurement of the domain size at early times after a spinodal decomposition could be an indirect way to determine the order of magnitude of $\kappa$ from experimental data. To our knowledge, no such measurement is available as of yet for lithium intercalated in graphite.

Another point linked to the value of $\kappa$ is the domain wall energy, which scales as the geometric mean of the gradient energy coefficient and the barrier height of the double-well potential in the free energy, $\Omega_a$. All of the interaction energies have been obtained by comparison of the model prediction to measurement of the equilibrium potential \cite{chandesris_thermodynamics_2019}, and are therefore physically realistic. As a consequence, the use of a larger value of $\kappa$ also implies that the energies of domain walls between domains and columns are larger than the physically correct value. We have not investigated the impact of this fact on the coarsening dynamics. Anyway, our present simulations of coarsening are limited to small systems (six layers), and to two dimensions (one-dimensional galleries). It would be interesting but computationally costly to perform simulations in larger system and in three dimensions in order to assess the coarsening dynamics more precisely.

\section{Conclusion}
In this paper, the dynamics of spinodal decomposition and domain coarsening in lithium-intercalated graphite after a quench (a sudden change in temperature) has been investigated, both by an analytical linear stability analysis and by numerical simulations of a multi-layer Cahn-Hilliard model. The multi-layer framework is needed to capture the well-documented phenomenon of staging, which is caused by repulsive inter-layer interactions between lithium atoms.

A normal-mode analysis of the linearized evolution equations has allowed us to follow separately the growth dynamics of each stage, and to compute analytically the associated growth rates. A stability spectrum that is typical for spinodal decomposition has been obtained for each stage, in the form of a quartic polynomial in the wave number for plane waves along the graphite planes. Beyond the linear regime, the decomposition in normal modes, corresponding to a discrete Fourier transform, remains useful to analyze the further evolution of the stage distribution, which can only be obtained by numerical simulations of the full evolution equations.

Our analysis predicts that stage 2 is always the fastest to grow after a quench, even for average compositions for which stage 3 is the thermodynamically stable state. Stage 3 emerges only at later times, during the nonlinear coarsening of stage 2, and progressively fills the entire system. While quench experiments are difficult to perform, our findings also yield a prediction for the kinetics of stage formation during the charging of a graphite particle: at the particle surface, stage 2 should always appear first. In contrast, if the equilibrium sequence of stages is followed, stage 3 (or higher) should appear first. How this competition between stages at the surface depends on the charge rate remains to be investigated in more detail.
Further work could also focus on the introduction, in the stage stability analysis, of electrochemical reactions at the particle surface~\cite{Bazant2017,Bazant2023,Agrawal2023}. This is necessary to understand if insertion reactions can impact the stability of the different stages and therefore modify the appearance of the different stages during lithiation/delithiation processes.

\section*{Acknowledgements}
The CEA Battery FOCUS program on "Multi-scale simulation of batteries applied to electrode materials" is acknowledged for the funding of the PhD program of Antoine Cordoba. 
%

\end{document}